\documentclass[preprint,aps]{revtex4}

\usepackage{graphicx}
\usepackage{dcolumn}
\usepackage{bm}
\usepackage{amsmath}

\newcommand{\rme}{\mathrm{e}}
\newcommand{\rmi}{\mathrm{i}}
\begin{document}

\title{Explosive growth of inhomogeneities
in the distribution of droplets in a turbulent air}
\author{S.A. Derevyanko$^{\#,\dag}$, G. Falkovich$^*$, K. Turitsyn$^{\&}$
and S. Turitsyn$^{\#}$}
\address{$^{\#}$Photonics Research Group, Aston University,
Birmingham B4 7ET, UK\\$^*$Physics of Complex Systems, Weizmann
Institute of Science, Rehovot 76100 Israel\\ $^{\&}$Landau
Institute for Theoretical Physics, Moscow 117940, Russian Federation \\
$^\dag$ Institute for Radiophysics and Electronics, Kharkov 61085,
Ukraine (on leave)}


\begin{abstract}
We study how the spatial distribution of inertial particles
evolves with time in a random flow. We describe an explosive
appearance of caustics and show how they influence an exponential
growth of clusters due to smooth parts of the flow, leading in
particular to an exponential growth of the average distance
between particles.
\end{abstract}
 \maketitle

Random compressible flows generally have regions where
contractions accumulate and density grows. Infinitesimal elements
expand or contract exponentially which can be characterized by the
set of Lyapunov exponents. Since the sum of the exponents is
non-positive \cite{Ruelle97,BFF,FF,FGV}, density tends to a
singular multi-fractal set with moments growing exponentially.
Both the evolution and the final state of density in spatially
smooth random flows have been described recently within some
models \cite{BFF,FGV,BGH,KG}. The flow of inertial particles is
compressible even when the flow of ambient fluid is incompressible
\cite{Maxey} so particles participate in the fractalization and
have some of their concentration moments growing exponentially
\cite{BFF}. On the other hand, every time there is a negative
velocity gradient exceeding the inverse viscous response time of
particles, faster particles from behind catch slower ones creating
folds in distribution and caustics \cite{FFS,WM}. Such breakdowns
of distribution lead to finite-time singularities and explosive
growth of some density moments. The goal of the present paper is
to describe the statistical evolution of concentration from a
uniform one to a set of clusters and voids and, in particular, to
describe the role of of folds in this evolution.

Because of folds, the problem of inertial particles in a flow is
kinetic rather than hydrodynamic \cite{FFS,Bec}. Analytic approach
to a realistic kinetic description does not seem to be feasible
now. On the other hand, the significant progress of analytic
Lagrangian methods \cite{FGV} makes it tempting to use them: to
follow, for instance, a couple of close particles and to account
only for a local velocity gradient. The question is: what can we
learn from the Lagrangian approach about the statistics of
particle concentration? To answer that, one needs a model that
allows to compare numerical  data from kinetics  with an analytic
Lagrangian solution. For that end we consider here the motion of
inertial particles in a one-dimensional random flow, which is
appropriate for our main goal to describe the role of breakdowns
that are one-dimensional in any space dimensionality. This model
is a subject of much interest from different perspectives
\cite{MW}.
Here we briefly review what is known and derive new results, in
particular, describe the statistics of the inter-particle
distances $R$. We also carry direct numerical simulation of
kinetics in this model and find the growth rates of the moments of
concentration $n$. It is only for smooth flows that one can
immediately convert $R$ into $n$ (in 1d simply taking $n=1/|R|$).
Since the flow of inertial particles has discontinuities, any
given interval between two chosen particles does not contain the
same particles all the time. Particles can enter and leave the
interval i.e. numerous folds appear in particle distributions
making nonlocal even the problem of describing single-point
density statistics. We show that indeed the growth rates of
density moments and inter-particle distances are different.

 Particle coordinate ${\bf q}$ and velocity ${\bf V}$ change
 according to
${d{\bf q}/ dt}= {\bf V}({\bf q}, t)$ and ${d{\bf V}/ dt}={[{\bf
u}({\bf q}, t)-{\bf V}]/\tau}$ with ${\bf q}({\bf r},0)={\bf r}$.
Here the viscous (response) time is $\tau=(2/9) ( \rho_0/\rho)
(a^2/\nu)$ with $a$ particle radius and $\rho_0,\rho$ particle and
fluid densities respectively. We treat the fluid velocity ${\bf
u}$ as a given random function of time and smooth function of
space coordinates.
Let us briefly remind some relevant properties of smooth
compressible random flows \cite{FGV}. The behavior of an
infinitesimal volume is governed by the local matrix of
derivatives (called strain matrix) taken in the Lagrangian frame
$s_{ik} =
\partial u_i/\partial x_k$. Considering the distance between two
fluid particles, ${\bf R}(t,{\bf r}_1-{\bf r}_2)={\bf q}({\bf
r}_1,t)-{\bf q}({\bf r}_2,t)$ one finds $\langle R^m\rangle\sim
\exp(E_mt)$ with  $E_m$ being a convex function of $m$. Density
can be expressed as $n(t) ={\rm det}^{-1}\,\partial R_i(t,{\bf
r})/\partial r_j$ (provided that the initial distribution is
uniform $n_0=1$) so that the Lagrangian moments $\langle
n^{-m}\rangle$ are related to space-averaged (Eulerian) moments
via $\langle n^{-m}\rangle =\langle n^{1-m}\rangle_E \sim
\exp(\Gamma_mt)$ (every trajectory comes with the weight
$n^{-1}$). Therefore, $\Gamma_{0}=0=\Gamma_1$ which correspond to
conservation of mass and volume (Lagrangian and Eulerian measures)
respectively. In one-dimensional (1d) smooth flows, $\Gamma_m=
E_{m}$.

In 1d,  one has for the distance $R(t)$ and velocity difference
$v(t)$ between two close inertial particles:
\begin{equation}\dot R={v}\,,\quad\tau \dot {
v}=s{  R}-{v}\qquad\Rightarrow\qquad\tau\ddot{R}+\dot{ R}=sR\
.\label{eq1}\end{equation} The substitution $R=\Psi\exp(-t/2\tau)$
turns (\ref{eq1}) into Schr\"odinger equation with a random
potential (Anderson localization), with space and the localization
length replacing time and the Lyapunov exponent.

The quantity $\sigma =v/R$ satisfies the Langevin equation driven
by the random noise $s(t)$
\begin{equation}
\dot \sigma=-\sigma ^{2} -\sigma /\tau +s/\tau \equiv
-dU/d\sigma+s/\tau \ .\label{1d1}
\end{equation}
Let us describe the probability of finite-time singularity
(explosion) $\sigma\to-\infty$ which corresponds to crossing of
particle trajectories. Such probability can be written as a path
integral over trajectories with
$\sigma(0)=\sigma_0,\sigma(T)=-\infty$:
\begin{eqnarray}P(T)\!=\!\!\int \!{\cal D}\sigma{\cal D}p{\cal
D}s \mathcal{P}\{s\}\exp\biggl\{\int_0^T\!\!ip\biggl[\dot\sigma+W-
\frac{s}{\tau}\biggr]dt'\biggr\}\,.\label{OF1}\end{eqnarray} Here
$\mathcal{P}\{s\}$ is the probability functional for $s$ and
$W=U'=(\sigma^2+\sigma/\tau)$. When $T$ is less than the average
time between explosions (defined below), $P(T)$ is given by the
single trajectory (optimal fluctuation  \cite{FL,CKLT,T5}) which
maximizes the probability and can be found by a saddle-point
integration of (\ref{OF1}).

First, consider $T$ which is much less than the correlation time
of the air gradient $s$. Then the optimal fluctuation corresponds
to $s=s_0$ which does not change during $t$. In this case the
integration over the processes $s(t)$ is reduced to the averaging
over a single value $s_0$ with the measure $P_s(s_0)$, which is a
single-time statistic of velocity gradient $s$. The saddle-point
integration over the fields $p,\sigma$ is reduced to solving the
equation (\ref{1d1}) with constant $s(t)=s_0$ and the boundary
conditions $\sigma(0)=\sigma_0,\sigma(T)=-\infty$. Straightforward
integration yields the following relation:
\begin{eqnarray}
 T &= &\tau\int_{\sigma_0}^{-\infty} \frac{d\sigma}{s_0-\sigma-\tau
 \sigma^2}\label{Teq}\\ &=&
 \tau(-1-4 s_0 \tau)^{-1/2}\left[{\pi-2 \arctan\left(\frac{1+2\sigma_0 \tau}{\sqrt{-1-4 s_0 \tau}}\right)}
 \right]\,,
\nonumber\end{eqnarray} which formally gives a relation between
the optimal value of $s_0$ and the collapse-time $T$. It is not
possible to find the analytic expression for $s_0(T)$ for a
general value of $\sigma_0$, however the situation greatly
simplifies for $\sigma_0=+\infty$. In this case the PDF $P(T)$ can
be interpreted as the distribution of time intervals between
consequent collapses. Note, that as long as the trajectory
starting from $\sigma_0=+\infty$ passes through all values of
$\sigma$ this distribution is also a lower estimate for the $P(T)$
for general $\sigma_0$. Substituting $\sigma_0=+\infty$ in
(\ref{Teq}) one obtains
\begin{equation}
 T = \frac{2\,\pi\tau}{\sqrt{-1 - 4 s_0 \tau}}
\end{equation}
or equivalently
\begin{equation}
 s_0 = -\frac{1}{4\tau}-\frac{\pi^2 \tau}{T^2}
\end{equation}
In this case the probability of collapse is given by
\begin{equation}P(T) = P_s(s_0) \left| {{ds_0}\over{dT}}\right|=
\frac{2\pi^2\tau}{T^3} P_s\left(-\frac{1}{4\tau}-\frac{\pi^2
\tau}{T^2}\right)\ .\label{short}\end{equation} One can see from
this expression that collapses occur only if there is a finite
probability of having sufficiently negative flow gradient, $s <
-1/4\tau$.
In particular for Gaussian gradients, $P_s(x) =
(\alpha/\pi)^{1/2}\exp(-\alpha x^2)$, the short-time asymptotics
is as follows: $P(T)\sim T^{-3}\exp(-\alpha\pi^2 \tau^2 /T^4 )$.

Consider now the case when the correlation time of $s$ is much
shorter than $T$. In this case, the noise can be effectively
considered as white Gaussian,  $\langle
s(t)s(0)\rangle=2D\tau^2\delta(t)$,  and
\begin{eqnarray}P(T)\!=\!\int {\cal D}\sigma\exp\biggl\{-\frac{1}{4D}
\int_0^T[\dot\sigma+W]^2dt'\biggr\}\ .\label{OF1}\end{eqnarray}
For $DT^3\ll1$, if follows from the saddle point approximation
that the probability is given by the optimal fluctuation (also
called "instanton" trajectory  \cite{FL,CKLT,T5}) which satisfies
$\ddot{\sigma} = W(\sigma) W'(\sigma)$ with the boundary
conditions $\sigma(0)=\sigma_0, \sigma(T) = -\infty$. After one
integration one obtains the following equation:
\begin{equation}
 \dot{\sigma} = -\sqrt{E + W^2}
\end{equation}
where $E$ is an integration constant, characterizing the trajectory.
This constant is determined by the boundary conditions:
\begin{equation} \label{Teq2}
 T = \int_{-\infty}^{\sigma_0} \frac{d\sigma}{\sqrt{E+W^2}}
\end{equation}
The probability of such fluctuation is given by $P(T) \sim
\exp(-A)$, where
\begin{equation} \label{Aeq}
 A = \int_0^T dt\frac{ (\dot{\sigma}+W)^2}{4D}  =
 \int_{-\infty}^{\sigma_0} \frac{d\sigma}{4D}\frac{(\sqrt{E+W^2}-W)^2}{\sqrt{E+W^2}}
\end{equation}
Unfortunately, the integrals (\ref{Teq2},\ref{Aeq}) can not be
expressed through known special functions, so we are able to get
analytical results only in some limiting cases. We will consider
the case $\sigma_0=+\infty$ following the same arguments as in the
preceding analysis. First, we consider the limit $E\tau^4\ll 1$
which as follows from (\ref{Teq2}) corresponds to large times $T
\sim \tau \log(1/E\tau^4) \gg \tau$. From the expression
(\ref{Aeq}) we have in the main order:
\begin{equation}
 A = \int_{-\infty}^\infty \frac{(|W|-W)^2d\sigma}{{4 D}|W|}
 =\int_{-1/\tau}^0  \!\!\!\frac {|W| d\sigma}{D} = \frac{1}{6D\tau^3}\,.
\end{equation}
We see, that in the main approximation the action does not depend
on the $T$, which has a simple interpretation: the collapses are
produced by universal tunneling processes, each having a
probability $\exp(-1/6D\tau^3)$ and characteristic time-scale
$\tau$. In order to find the $T$ dependence of the total
probability we should study the fluctuations around this instanton
\cite{AKOSW99} which would involve some bulky calculations.
However, for the intermediate region of $T\ll \tau
\exp(1/6D\tau^3)$ one can treat these tunnelings as a Poissonian
process and predict the linear behavior $P(T) \sim T/\tau
\exp(-1/6D\tau^3)$. This expression is certainly not true in the
case $D\tau^3 \lesssim 1$ when the action $A$ is not large and the
saddle-point approximation is not applicable. Another limiting
case, which can be studied analytically corresponds to the very
high "energies" $E\tau \gg 1$ where one can neglect the linear
$\sigma/\tau$ terms in (\ref{Teq2},\ref{Aeq}), so that one has
\begin{eqnarray}
 T = \frac{ \Gamma(1/4)^2}{2 \sqrt{\pi}E^{1/4}},\qquad
 A = \frac{\Gamma(1/4)^8}{96 \pi^2 D T^3 }\approx \frac{31.5}{D T^3}
\end{eqnarray}
The crossover between the two regimes happens at $T \sim \tau$. To
summarize, for the white $s(t)$ one gets
\begin{eqnarray}
P(T)\sim\left\{\begin{array}{ll}\exp(-c/DT^3), &T<\tau,\\
T \exp(-1/6D\tau^3)&\tau<T<
\tau\exp(1/6D\tau^3),\label{long}\end{array}\right.
\end{eqnarray}
where $c = {[\Gamma(1/4)]^8}/{96 \pi^2} \approx  {31.5}$.

Since we consider dilute distribution of particles and neglect
their pressure, then $\sigma$ changes sign after the explosion as
the fast particle overcomes the slow one. That is the flux of
probability that goes to $\sigma\to-\infty$ returns from
$\sigma\to+\infty$. That allows for the steady-state probability
density function (PDF) having constant probability $F$ flux equal
to the number of breakdowns per unit time. Such PDF must have
$P(\sigma)\approx F\sigma^{-2}$ at $\sigma\to\pm\infty$. If, as is
usually the case, the initial $P(\sigma,0)$ does not have power
tails, they appear at $t=+0$ according to  $P(t,\sigma) \propto
P(t)\sigma^{-2}$ and (\ref{short},\ref{long}).

When $\sigma\to-\infty$, $R\to0$. To establish the sufficient
condition for negative moments of the distances to blow-up in a
finite time, introduce $R_{l,k}=\langle \sigma^l R^{k}\rangle$.
Assuming even $k$, using (\ref{1d1}) and Cauchy inequality
$R_{1,k}\leq R_{2,k}^{1/2}R_{0,k}^{1/2}$ we get for
$Z=R_{0,k}^{1/k}$ the majoring inequality $k(Z_{tt}+Z_t/\tau
)\geq0$. For positive $k$, it means smooth evolution with $Z$
growing. For negative $k$, this inequality gives $Z(t)\leq
Z(t_{1}) +\tau Z_{t}(t_{1}) (1-e^{-(t-t_{1}) /\tau })$. This means
that $Z$ turns into zero, and respectively, the negative momenta
of the distances ($k < 0$) will blow up in a finite time if at
some $t_{1}$: $Z+\tau Z_{t}<0$ or in other terms,
${\tau dR_{0,k}/dt}>|k|R_{0,k}$. This
 condition is readily satisfied for most random processes $s(t)$,
the detailed analysis will be published elsewhere.

In the rest of the paper we approximate the flow gradient $s(t)$
in the particle reference frame by a white noise, which is
quantitatively good for heavy particles and give a qualitatively
correct description in other cases. In the white case, a variety
of analytic results can be obtained, some translated from the
localization theory and super-symmetric quantum mechanics
\cite{Halp,AKOSW99} and some original that we derive here. The
steady-state PDF can be found explicitly \cite{Halp}
\begin{eqnarray}&&\!\!\!\!\!\!\!\!\!\!\!\!
P_0=\frac{F}{D}\exp\!\left[-\frac{U(\sigma)}{D}\right]
\int\limits_{-\infty}^{\sigma}\!\!\!
\exp\!\left[\frac{U(\sigma')}{D}\right] d\sigma'\,,\label{1d3}
\end{eqnarray}
with the flux  $F= D {\partial P_0}/{\partial \sigma}+
\left(\sigma^2+ {\sigma}/{\tau}\right)P_0 \approx
(2\pi\tau)^{-1}\exp[-1/(6D\tau^3)]$ for $D\tau^3 \ll1$ (the
dimensionless Stokes number $D\tau^3=St$  measures the inertia of
the particle). At $St\gg1$, $F\approx 0.2D^{1/3}$ \cite{MW}, note
that the average time between breakdowns is much smaller than
$\tau$ in this limit. The Lyapunov exponent $\langle\sigma\rangle$
changes sign at $St_*\approx 0.827$ \cite{MW}:
$\langle\sigma\rangle\approx -D\tau^2/2$  at $St\ll St_*$ and
$\langle\sigma\rangle\sim D^{1/3}$ at $St\gg St_*$. That means
that small particles cluster while large ones mix uniformly.

Note that the Gibbs state $\exp(-U/D)$ is non-normalizable in this
case. The flux state (\ref{1d3}) minimizes entropy production
\cite{FT}. It can be shown that it is indeed the asymptotic
solution at $t\to\infty$ \cite{Gaw}.

To describe the  joint statistics of $\sigma$ and $R$
we introduce the generating function $ Z_{k}(\sigma ,t)=\langle
\delta \left[ \sigma \left( t\right) -\sigma \right]
R^k(t)\rangle$, which satisfies the equation
\begin{equation}
\frac{\partial Z_{k}}{\partial t}=k\sigma Z_{k}+\frac{\partial
}{\partial
\sigma }\left( \frac{\sigma }{\tau }+\sigma ^{2}+{D}\frac{\partial }{%
\partial \sigma }\right) Z_{k}\,.\label{gen1}
\end{equation}
Substitution $Z_{k}=\Psi (\sigma ,t)\exp[ -U/2D]$  turns it into
the Schr\"odinger equation in a double well, which has been a
subject of numerous works related to tunnelling and instantons
(see e.g. \cite{Bog,Wit,Turb,AKOSW99}. Following
\cite{Turb,AKOSW99} we first find (non-normalizable) solutions
$\exp(\gamma_kt/\tau-U/D)f_k(\sigma)$ with $f_k$ being polynomials
and then the conjugated solutions by the method of variable
constants. For example, there are steady states $Z_0=P_0$
 and $$Z_1(\sigma)=
(1+\sigma\tau)\exp\!\left[\frac{U(\sigma)}{D}\right]
\int_{-\infty}^{\sigma}\!\!\!\!\!
\exp\!\left[\frac{U(\sigma')}{D}\right]{d\sigma'\over(1+\sigma'\tau)^2}\
.$$ In particular, this solution allows one to obtain the mean
velocity difference between two particles at the distance $a$:
$a\int\sigma Z_1(\sigma,t)\,dt$ needed, for instance, to calculate
the collision rate. The growth rates of the moments of
inter-particle distance can be obtained from (\ref{gen1}) or in a
straightforward way by writing
\begin{equation}
\dot R_{l,k}=-lR_{l,k}/\tau-(l-k)R_{l+1,k}+l(l-1)DR_{l-2,k}\
,\label{moments}\end{equation} where the higher moments are
expressed only via lower ones. Assuming that for a given $k$ all
$R_{l,k}\propto \exp(\gamma_k t)$ we get for $\gamma_k$ the
$(k+1)\,$-st order algebraic equation. For the second moment  one
gets $\gamma_{2} \left( \gamma_{2} +\tau ^{-1}\right) \left(
\gamma_{2} +2\tau ^{-1}\right) -4D=0$ which gives $\gamma_{2}
\approx 2D\tau ^{2}$ for $D\tau ^{3}\ll 1$ and $\gamma_{2} \approx
\left( 4D\right) ^{1/3}$ for $D\tau ^{3}\gg 1$.

For arbitrary $k$, we find asymptotics. If $D\tau^3k^2\ll1$ then
$\gamma_{k}\approx D\tau^2k(k-1)/2$. When $D\tau^3k^2\gg1$, the
determinant of (\ref{moments}) is approximately
$\gamma_{k}^{k+1}-\gamma_{k}^{k-2}Dk(k-1)\sum$, where
$\sum=\sum_1^{k}i(k-i)\propto k^2$ and $\gamma_{k}\propto
(Dk^4)^{1/3}$. Let us compare the growth rates of the distance
moments for the inertial particles with those for smooth
compressible short-correlated flow. For the latter, $\gamma_k\sim
k(k-1)$ while for the former the dependence is parabolic only for
low-order moments in the low-inertia limit $D\tau^3k^2\ll1$. High
moments correspond to high inertia and have $\gamma_k\sim
(Dk^4)^{1/3}$ even for $St\ll1$. Note that conservation requires
$\gamma_0=\gamma_1=0$  for inertial particles as well.

\begin{figure}[h!]
\centerline{\includegraphics[width=5.1cm,height=3.4cm]{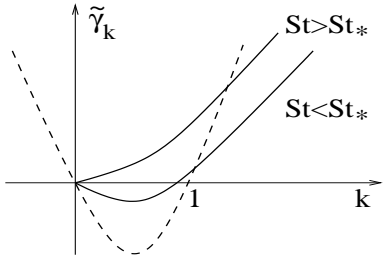}}
\caption{Growth rates of distance moments for a smooth flow
(broken line) and inertial particles  for different Stokes numbers
(two solid lines).} \label{fig:0}
\end{figure}
Since $R$ is sign-changing for inertial particles, the statistics
of $|R|$ deserves separate study, particularly for comparison with
the concentration. The equation for the time derivative of $\tilde
R_{lk}=\langle \sigma^l |R|^{k}\rangle$ differs from
(\ref{moments}) by the extra term $2\langle
\sigma^{l+1}R^{k+1}\delta(R)\rangle$, which is nonzero for $l=k$.
As a result, the growth rates $\tilde\gamma_k\equiv\tilde
R_{lk}^{-1} d\tilde R_{lk}/dt$ differ remarkably from $\gamma_k$.
The most dramatic new effect can be readily appreciated since
$\tilde\gamma_k$ are related to the Lyapunov exponent via
$\langle\sigma\rangle= ({d\tilde\gamma_k/d k} )_{k=0}$. At high
inertia, when $St>St_*$ and $\langle\sigma\rangle$ is positive, it
is thus evident that $\tilde\gamma_1>0$ as seen from the sketch in
Fig.~\ref{fig:0}. Nonzero growth rate of $\langle |R|\rangle$ is a
remarkable qualitatively new effect with a clear physical meaning:
in every breakdown, extra particles enter the interval between the
two particles that we follow; the interval length must grow to
ensure conservation of the total number of particles. From this
interpretation, it is clear that the growth rate must be nonzero
at low inertia as well, when it must be proportional to the
exponentially small rate of explosions: $\gamma_1\sim F\sim\exp (-
{1}/{6 St})$. Remarkably, one can also establish asymptotically
exact pre-exponential factor. Consider ${d|R|}/{dt} = \sigma |R| +
2 \delta(R)\sigma R^2$. The growth of $\langle|R|\rangle$ must be
determined by the last term, which accounts for the breakdown
processes, since $\langle R\rangle$ does not grow. In order to
obtain the explicit expression for $\tilde \gamma_1$ we first
analyze the dynamical equation on the stages between the
breakdowns, which formally coincides for both $R$ and $|R|$ and
then account for breakdowns explicitly. For delta-correlated $s$,
we can break the time interval into pieces with independent
evolution (markovian
property). Using this fact and multiplicative nature of 
(\ref{eq1}) one can derive the following identity
\begin{equation} \label{prod}
 \left\langle{|R(t)|\over|R(0)|}\right\rangle =
 \left\langle\frac{|v(t_1)|}{|R(0)|}\right\rangle
 \left\langle\frac{|R(t)|}{|v(t_N)|}
 \right\rangle
 \prod_{k=2}^N \left\langle\frac{|v(t_k)|}{|v(t_{k-1})|}\right\rangle
\end{equation}
Here $t_1..t_N$ are the times when the breakdowns happened, $v(t_k)$
are absolute values of the velocities in these breakdowns. All the
averages in this expression correspond to the
dynamics between the breakdowns, for which 
(\ref{eq1}) can be solved explicitly. In order to study the dynamics
between breakdowns we introduce the function $Z(\sigma,\sigma_0,t)$
which is the solution of (\ref{gen1}) with $k=1$ and the initial
condition $Z(\sigma,\sigma_0,0) = \delta(\sigma-\sigma_0)$. In
contrast to the previous analysis we consider different boundary
conditions for the function $Z(\sigma,\sigma_0,t)$. Namely we will
assume that there is no flux at $\sigma=+\infty$, which means that
$Z(\sigma,\sigma_0,t)$ decays exponentially there. In this case
$Z(\sigma,\sigma_0,t)$ can be interpreted as  $Z(\sigma,\sigma_0,t)
= \overline{R(t)/R(0)}$ where the averaging is performed only on the
trajectories which had no breakdown up to time $t$ and which satisfy
the following boundary conditions
$\sigma(t)=\sigma,\sigma(0)=\sigma_0$. Note, that for such
trajectories we have $|R(t)/R(0)|=R(t)/R(0)$. We are able to fix the
breakdown moment at $t$ by taking the limit $\sigma\to -\infty$.
Analogously the limit $\sigma_0\to +\infty$ fixes the preceding
breakdown moment at $t=0$. In order to analyze the product
(\ref{prod}) we introduce three new functions:
\begin{equation}
 \frac{|v(t_k)|}{|R(t_k-t)|}= J_{+}(\sigma_0,t) = -\sigma^3 Z(\sigma,\sigma_0,t)|_{\sigma\to-\infty}
\end{equation}
\begin{equation}
 \frac{|R(t_k+t)|}{|v(t_k)|}= J_{-}(\sigma,t) = \sigma_0^{-1} Z(\sigma,\sigma_0,t)|_{\sigma_0\to\infty}
\end{equation}
\begin{equation}\label{mdef}
 \frac{|v(t_{k+1})|}{|v(t_k)|}= M(t_{k+1}-t_k) = \sigma_0^{-1}J_{+}(\sigma_0,t_{k+1}-t_k)|_{\sigma_0\to\infty}
\end{equation}
These function have the following meaning. $J_+(\sigma_0,t)dt$ is an
average ratio of $|v(t)/R(0)|$ for trajectories with the boundary
condition $\sigma(0)=\sigma_0$ which had breakdown at the interval
$(t,t+dt)$ . Analogously $J_-(\sigma,t)$ is a ratio of $|R(t)/v(0)|$
for the trajectories which emerged after breakdown at $t=0$.
Finally, $M(t)dt$ gives us the average ratio of the velocities
$v(t)/v(0)$ between the two breakdowns which happened at time $0$
and in the interval $(t,t+dt)$. Note that the normalization factor
$\sigma^{-3}$ in the definition of $J_{+}(\sigma_0,t)$ accounts for
the flow of trajectories  given by $(\sigma^2+\sigma \tau^{-1}) Z$.
With such a normalization one has $\int_\Omega dt J_{+}(\sigma_0,t)$
is the average ratio of $|v(t_k)/R(0)|$ of all trajectories which
satisfy $\sigma(0)=\sigma_0$ and had a single breakdown at time $t_k
\in \Omega$. After introducing these three new functions we are able
to average the ratio $|R(t)/R(0)|$ over all trajectories with an
arbitrary number of breakdowns happened at all possible times. We
can write formally
\begin{equation}
\left\langle{|R(t)|\over|R(0)|}\right\rangle = \int d\sigma
\left[Z(\sigma,\sigma_0,t)+\sum_{N=1}^\infty \int \prod_{k=1}^N d
t_k
J_{+}(\sigma_0,t_1)J_{-}(\sigma,t_k)\prod_{j=1}^{N-1}M(t_{j+1}-t_j)\right]
\end{equation}
This expression can be simplified by turning to the Laplace
transform representation:
\begin{equation}
\left\langle \frac{|R(t)|}{|R(0)|}\right\rangle = \int \frac{d\sigma
ds}{2\pi i}\exp(s t)\left\{Z^s+J_+^sJ_-^s+J_+^s M^s J_+^s+
\dots\right\}= \int \frac{d\sigma ds}{2 \pi i}\exp(s t)\left\{Z^s
+\frac{J_+^s J_-^s}{1-M^s}\right\}\label{RReq}
\end{equation}
Where the upper index $s$ corresponds to the Laplace transform of a
function:
\begin{equation}
 F^s = \int_0^\infty dt \exp(-s t) F(t)
\end{equation}
Long time asymptotic of both expressions is determined by the most
left pole or other singularity of the integrated functions. One can
easily note all three functions $Z^s,M^s$ and $J_{\pm}^s$ have the
same poles $s=E_k$. Therefore the long time asymptotic is determined
either by $E_0$ or the most left solution of the equation $M^s=1$.
We will show that in our cases the asymptotic is indeed determined
by the later singularities. First we want to show that $M^0=-1$. The
Laplace transform of $Z$ obeys the following equation:
\begin{equation}
 \left[s -\partial_\sigma \left(\frac{\sigma}{\tau}+\sigma^2\right)
 -D\partial_\sigma^2-\sigma\right] Z_\sigma = \delta(\sigma-\sigma_0)
\end{equation}
here we have inserted the initial conditions
$Z(\sigma,t=0)=\delta(\sigma-\sigma_0)$ in the r.h.s. explicitly.
Remarkably this equation may be rewritten in the divergent form
after the substition $Z = (\tau^{-1}+\sigma)^{-1} \Pi$. This fact
was probably first noted in \cite{BY86}. New equation acquires the
following form:
\begin{equation}\label{fp}
 \left[s+\partial_\sigma\left\{-\sigma(\tau^{-1}+\sigma)+\frac{D}{\tau^{-1}+\sigma}\right\}
 -D\partial_\sigma^2\right]\Pi=(1+\sigma_0)\delta(\sigma-\sigma_0)
\end{equation}
In order to find $\Pi^0$ we have to set $s=0$ in this equation,
after which it can be easily integrated:
\begin{equation}\label{z0eq}
 Z^0(\sigma,\sigma_0) = \frac{1}{D}\frac{\tau^{-1}+\sigma_0}{\tau^{-1}+\sigma}U(\sigma)\int_{-\infty}^{\sigma'}
 U^{-1}(\sigma_1) d \sigma_1
\end{equation}
where $\sigma'=\min(\sigma,\sigma_0)$ and
$U(\sigma)=(\tau^{-1}+\sigma)^2\exp(-\sigma^2/2D\tau-\sigma^3/3 D)$.
Straightforward integration of (\ref{z0eq}) yields the expected
result:
\begin{equation}
 M^0 = \lim \left(-\frac{\sigma^3}{\sigma_0}\right)
 Z^0(\sigma,\sigma_0) = -1
\end{equation}
where we assumed the limit $\sigma\to-\infty,\sigma_0\to +\infty$.
Returning now to the determination of long time asymptotic of
$|R(\sigma,t)|$ we conclude that it is given by the solution of the
equation $M^s = 1$. In general case the explicit form of $M^s$ can
not be found analytically, so in the next consideration we will
assume the limit $D\tau^3\ll 1$. In this case we know that the
growth rate of $|R|$ is parametrically small, so that the solution
$s$ is almost zero. So, we need to know what is the behaviour of
$M^s$ near zero. In order to analyze it we turn the equation
(\ref{fp}). After the substitution $\Pi=U^{1/2}(\sigma)\Psi$ we
arrive to the quantum-mechanical problem
\begin{equation}
 \hat{H}\Psi = -s \Psi,\qquad \hat{H} =
 -D\partial_\sigma^2 +
 \frac{\sigma^2(\tau^{-1}+\sigma)^2}{4D}-\frac{1}{2\tau}-2\sigma
\end{equation}
Such asymmetric double-well Hamiltonians have been extensively
studied in the literature, see e.g. \cite{BY86,AKOSW99} where the
spectrum of $\hat{H}$ was analyzed in the limit $D\to 0$. Omitting
the details we will just note that in the main order the ground
state energy of $\hat{H}$ is negative, with absolute value $E_0 = -E
= - (D\tau^2/2\pi) \exp(-1/6D\tau^3)$ while the other energy levels
are positive and are of order unity (are not exponentially small).
From the hermiticity of $\hat{H}$ it follows that the general form
of $M^s$ will be the following:
\begin{equation}
 M^s = \sum_k \frac{c_k}{s+E_k}
\end{equation}
where $c_k=-\sigma^2
U^{1/2}(\sigma)U^{-1/2}(\sigma_0)\Psi_k(\sigma)\Psi_k(\sigma_0)$
taken at limits $\sigma\to-\infty,\sigma_0\to +\infty$. All $c_k$ in
the main order are proportional to $c_k \propto \exp(-1/6D\tau^3)$
and are thus exponentially small, while out of all energies $E_k$
only $E_0$ is exponentially small. Therefore in the vicinity of
$s=0$ only the term with $k=0$ is relevant. Although we could find
the $c_0$ {\it ab initio} we won't do that and will instead use the
fact that $M^0=-1$, which immediately yields $c_0=E$. Therefore in
order to find the expression for the growth rate of $|R|$ we have to
solve the algebraic equation $E/(s-E)=1$ from which we finally
obtain the growth rate exponent of $|R|$.
\begin{equation}\tilde{\gamma}_1\tau=s\tau=2E\tau=({St}/{\pi})\,\exp (- {1}/{6
St})\ .\label{pred1}\end{equation} This final expression shows that
indeed the leading singularity in (\ref{RReq}) is determined by the
solution of $M^s = 1$. The growth rate $\tilde{\gamma}_1$ is
exponentially small because it is determined by the rare breakdown
events. Let us emphasize that we have established asymptotically
exact pre-exponential factor in (\ref{pred1}).

\begin{figure}[h!]
\centerline{\scalebox{1.0}{\includegraphics[bb= 0 0 246
186]{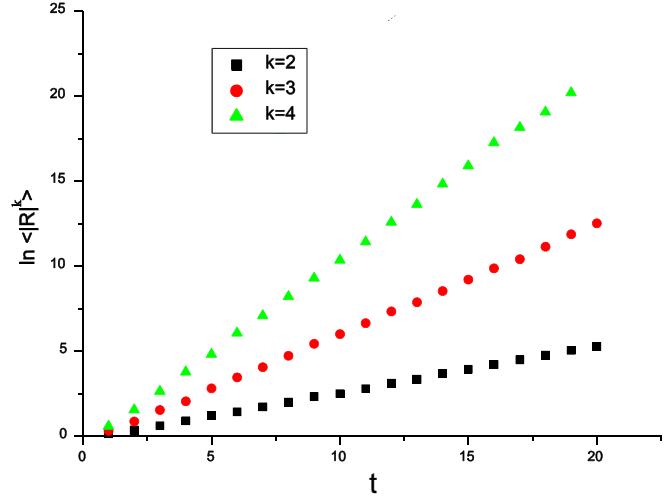}}} \caption{The moments  $<|R|^{k}>$ for
$k=2,3,4$ for $St=0.2$. Time is normalized by $\tau$.}
\label{fig:1}
\end{figure}
\begin{figure}[h!]
\centerline{\scalebox{1.0}{\includegraphics[bb= 0 0 243
176]{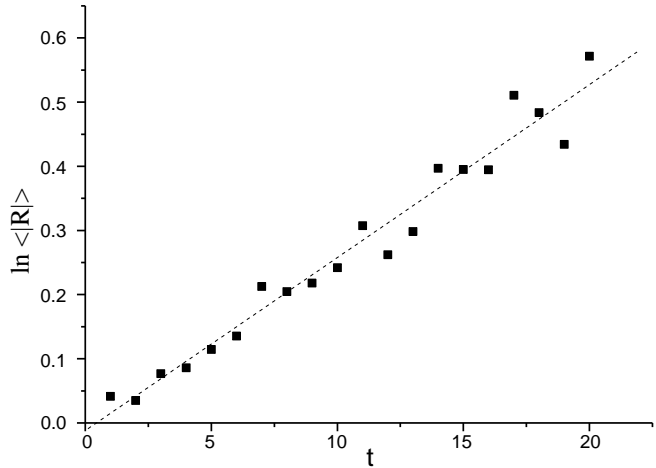}}} \caption{Because of inertia the modulus of
particle separation $<|R|>$ grows ($St=0.2$). } \label{fig:1b}
\end{figure}

We now present the results of numerical simulations of the growth
of particle separation $<|R|^{k}>$ in Lagrangian frame and of
negative moments of density $<n^{-k}>$ in Eulerian frame. The
method used to obtain the growth rates is the Multicanonical Monte
Carlo \cite{BergBook}, a technique of adaptive importance sampling
which boosts the probability of rare events that determine large
negative moments. The Lagrangian results  were obtained solving
(\ref{eq1}). The results presented in
Figs.\ref{fig:1},\ref{fig:1b} confirm an exponential growth of
$\langle|R|^k \rangle$.
\begin{figure}[h!]
\centerline{\scalebox{1.0}{\includegraphics[bb= 136 118 366
233]{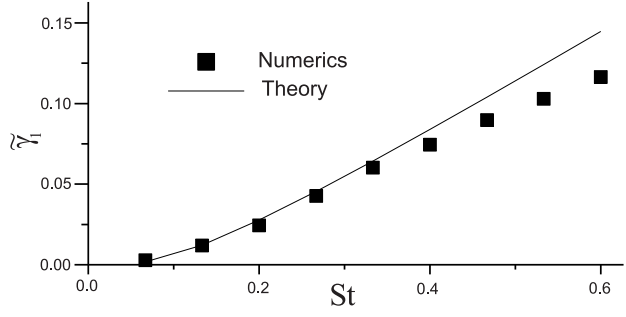}}} \caption{The growth rate $\tilde{\gamma}_{1}$ vs
the Stokes number. The solid curve represents theoretical
prediction.} \label{fig:2}
\end{figure}

We also observe an exponential growth of the particle separation,
$<|R|>$. Figure \ref{fig:2} shows a good agreement between the
numerics and the theoretical prediction (\ref{pred1}) up to a
fairly large $St\simeq0.35$.

For the 1D Eulerian simulations the density field is given by the
following expression
\begin{equation}
n(x,t)=\int d x_0 n_0(x_0) \delta(x(t|x_0)-x) \label{Euler-n}
\end{equation}
where $n_0(x_0)$ is an initial Eulerian density distribution
(which we assume uniform) and $x(t|x_0)$ is a Lagrangian
trajectory of a particle.

This trajectory is obtained from the system of the ODEs
(characteristic equations):
\begin{eqnarray}
\frac{d}{d t} \, x(t|x_0)&=&v(t), \quad x(0|x_0)=x_0, \label{eq-x}
\\
\frac{d}{dt} \, v(t) &=& -\frac{v(t)-u(x(t|x_0),t)}{\tau}
\label{eq-v}
\end{eqnarray}
where $u(x,t)$ is the Eulerian Gaussian velocity of the turbulent
flow.

We assume that it is delta-correlated in time and has a spatial
correlation length $l_c$:
\begin{equation}
<u(x,t)u(x',t')>= B(x-x')\delta(t-t'), \quad B(x)=B_0
\mathrm{e}^{-x^2/l^2_c} \label{correlation}
\end{equation}
The specific form of the correlation function $B$ is not
important.  Eulerian field $u(x,t)$ is related to the Lagrangian
process $s(t)$ (see Eq. (\ref{1d1})) via $s(t)=
\partial u(x(t|x_0),t)/\partial x$. From this it follows that
$St\equiv D \tau^3 =(\tau/2)\, |B''(0)|=(\tau/l_c^2)\, B_0 $.
Prior to solving system of ODEs (\ref{eq-x}), (\ref{eq-v}) one has
to generate 1D Eulerian velocity field $u(x,t)$ with the
prescribed correlation function (\ref{correlation}). The algorithm
for this is fairly standard (See. e.g. \cite{MajdaKramer}). First
we notice that since the field $u(x,t)$ is delta correlated in
time its temporal regularization is trivial. Introducing discrete
temporal step $\Delta t$ at each time step, $n$, we now need to
generate spatially distributed Gaussain field $u_n(x)$ with the
correlation property $<u_n(x)u_m(x')>= B(x-x') \, \delta_{mn}$. In
order to generate the field $u_n(x)$ we utilise the Fourier
method. Indeed the field $u_n(x)$ can be represented as a
following Fourier integral
\begin{equation}
u_n(x)=\intop^{\infty}_{0} \, \cos(2\pi k x)\, [2E(k)]^{1/2}
\xi_n(k) dk + \intop^{\infty}_{0} \, \sin(2\pi k x)\,
[2E(k)]^{1/2} \eta_n(k) dk \label{u-Fourier}
\end{equation}
where $\xi_n(k)$ and $\eta_n(k)$ are independent real Gaussian
processes with the following properties:
\begin{equation}
\begin{split}
\langle \xi_n(k) \rangle&=\langle \eta_n(k) \rangle=0\\
<\xi_n(k)\xi_m(k')>&= \langle \eta_n(k) \eta_m(k') \rangle =
\delta(k-k')\,\delta_{mn}
\end{split}
\label{xi-correl}
\end{equation}
and $E(k)$ is an energy spectrum of the random field $u_n$, it
coincides with the Fourier transform of the correlation function
$B(x)$:
\begin{equation}
E(k)=\intop^{\infty}_{-\infty}  \rme^{2\pi \rmi k x} B(x) dx=
B_0\,\sqrt{\pi l_c^2} \exp\left[-\pi^2 l_c^2 k^2\right]
\label{spectrum}
\end{equation}
We then use a discrete version of (\ref{u-Fourier}):
\begin{equation}
u_n(x)  \approx  \sqrt{\,E(0)\,\Delta k}\,\xi_0^n +\sum_{j=1}^M \,
\sqrt{2\,E(k_j)\,\Delta k}\, \left[\xi_j^n \cos(2\pi k_j \, x) +
\eta_j^n \sin(2\pi k_j \,x) \right] \label{u-Fourier-discr}
\end{equation}
Here we have partitioned the Fourier space into $M$ intervals, so
that the wavevectors $k_j=j\Delta k$ denote the locations of the
equispaced grid points. Variables $\xi_j^n$ and $\eta_j^n$ form a
set of independent standard Gaussian variables (mean zero and unit
variance). Because of the nature of the Fourier method the
synthetically generated field $u_n(x)$ will contain an intrinsic
spatial period $\lambda_F=(\Delta k)^{-1}$. Naturally one wishes
to make it much bigger than the characteristic scale of the system
$L$. On the other hand one has to ensure that all we have enough
harmonics in (\ref{u-Fourier-discr}) to sample the peak of the
function $E(k)$. These two requirements can be met assuming  $
(l_c M)^{-1} \lesssim \Delta k \ll L^{-1}$.
\begin{figure}[b!]
\centerline{\scalebox{1.0}{\includegraphics[bb= 122 156 366
286]{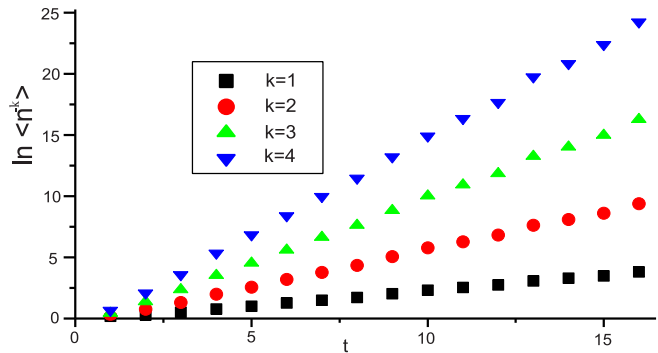}}} \caption{The Eulerian moments $<n^{-k}>$ for
$St=0.2$.} \label{fig:3}
\end{figure}
\begin{figure*}
\centerline{\scalebox{0.9}{\includegraphics{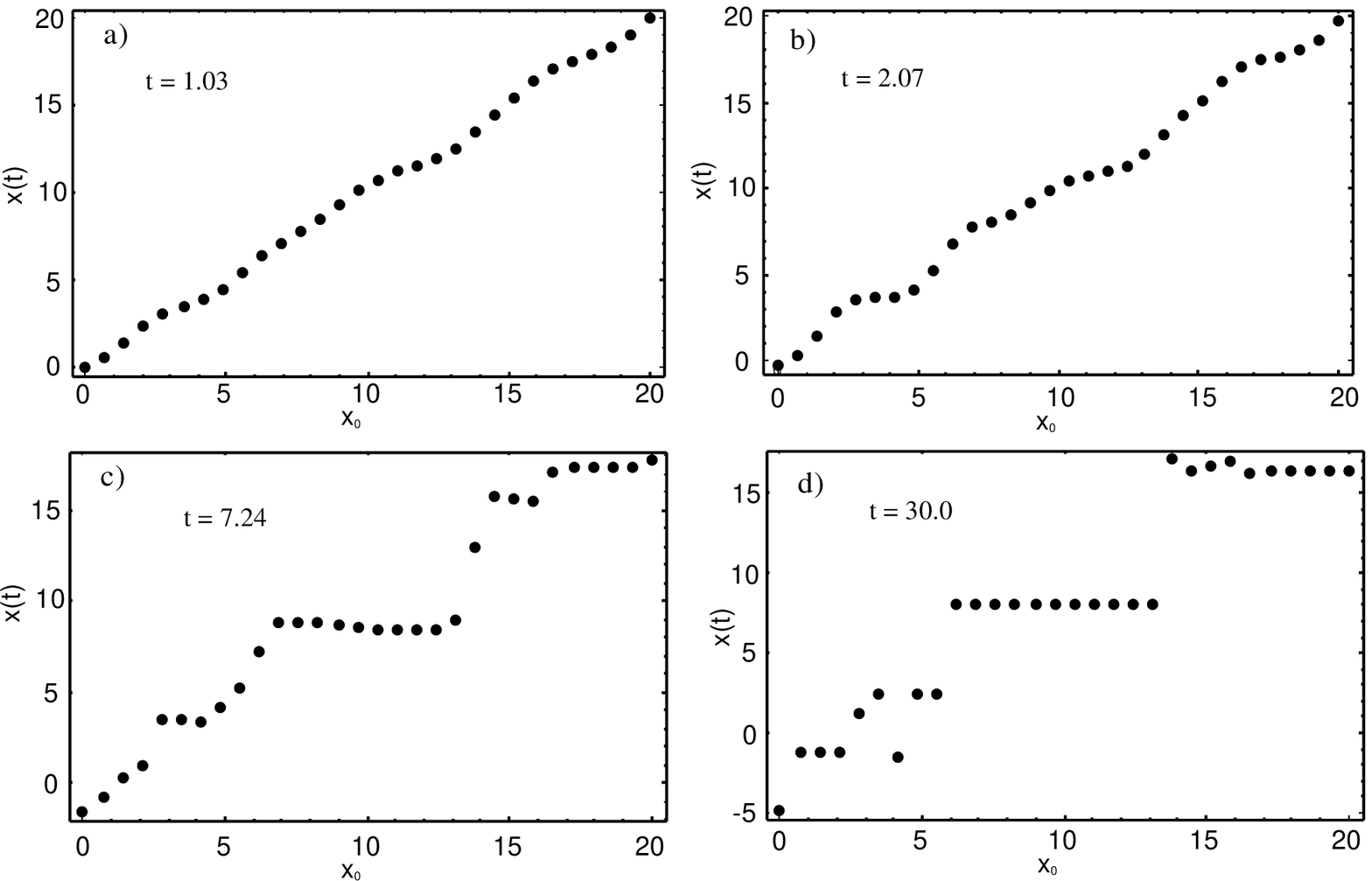}}}
\caption{The evolution of 30 Lagrangian markers with time. The
time is normalized to $\tau$ and increases from a) to d).
$St=0.4$} \label{fig:4}
\end{figure*}

Once we generated the synthetic Eulerian velocity field $u(x,t)$
we use a method of Lagrangian markers to obtain the Eulerian
particle density at each point (using effectively formula
(\ref{Euler-n})). We introduce a chain of $N_L$ representative
Lagrangian markers connected by some fictitious ``strings''.  Each
``string'' contains a large constant number of uniformly
distributed real particles. This number is fixed for each string,
it does not change during the evolution and is determined by the
initial density distribution. During the evolution, the strings
deform according to the Lagrangian dynamics of the initial
markers. In particular the occurrence of explosions in Lagrangian
frame corresponds to the formation of \textit{folds} in the chain
of markers. In order to obtain numerically the local Eulerian
particle density at a given point we count the number of strings
passing through this point and then for each string determine the
contribution to the density as a ratio $N_i/l_i$ where $N_i$ is
the number of particles in the string and $l_i$ is the current
length of the string. In Fig. \ref{fig:3} we plot the first four
negative moments of $n$. Similarly to Lagrangian moments, Eulerian
moments also grow exponentially: $<n^{-k+1}> \propto \exp(\Gamma_k
t)$. The table compares $\Gamma_k$ and Lagrangian
$\tilde{\gamma}_k$ given by (\ref{moments}) for $St=0.1$ and
$St=0.2$.
We see that Lagrangian breakdowns (Eulerian folds) violate
$\Gamma_k=\tilde{\gamma}_k$ that one would have for a smooth flow.
We do not have a meaningful parametrization for the dependencies
of $\tilde{\gamma}_k-\Gamma_k$ on $k$ and $St$.
It is likely that rare explosions cannot be completely
disentangled from the exponential evolution.

\begin{table}[h!]
\begin{tabular}{|c|c|c|c||c|c|c|c|}
\hline $k$ & $\tilde{\gamma}_k$ & $\Gamma_k$ &
$\tilde{\gamma}_k-\Gamma_k$ & $k$ &
$\tilde{\gamma}_k$ & $\Gamma_k$ & $\tilde{\gamma}_k-\Gamma_k$\\
\hline 1 & $0.006$ & ----- & ----- &1 & $0.028$ & ----- & -----\\
\hline 2 & $0.158$ & $0.146$  & $0.012 \pm 0.003$ & 2 & $0.274$
 & $0.250$ & $0.025 \pm 0.002$\\
\hline 3 & $ 0.393$ & $0.374$ & $0.019 \pm 0.005$ & 3 & $0.643$
 & $0.611$ & $0.032 \pm 0.005$ \\
\hline 4 & $0.695$ & $0.666$ & $0.029 \pm 0.006$ & 4 & $1.098$
& $1.054$ & $0.044 \pm 0.008$ \\
\hline 5 & $1.054$ & $1.012$  & $0.043 \pm 0.009$ & 5 & $1.627$
 & $1.564$ & $0.063 \pm 0.009$ \\
\hline 6 & $1.459$ & $1.403$ & $0.056 \pm 0.010$ & 6 &
$2.223$ & $2.131$ & $ 0.098 \pm 0.012$\\
\hline
\end{tabular}
\caption{The comparison of Eulerian and Lagrangian growth rates
for $St=0.1$ (left) and $St=0.2$ (right).} \label{tab1}
\end{table}

In 1D case there is a very simple way of visualizing the dynamics
of the caustics. At a given time moment, $t$, one can plot the
final displacement of a particle, $x(t)$ versus its initial
position $x_0$. The Eulerian density distribution can be obtained
by projecting the plot onto the coordinate axis $x(t)$. At the
time moment $t=0$ the curve is just a straight line at the half
the right angle to the axes. During the evolution it will deform,
according to the Lagrangian dynamics of individual particles
(\ref{eq-x}),(\ref{eq-v}) eventually leading to the formation of
folds illustrating the nonlocal nature of Eulerian density.

In Fig.\ref{fig:4} we plot three stages of evolution of particle
distribution. We take $N_L=30$ initially equispaced Lagrangian
markers and follow the evolution of the function $x(x_0)$ through
time for a particular realization of the velocity field. We
observe that at the initial stage (Fig.\ref{fig:4}a) the particle
displacements are small so that the density distribution is smooth
and there is one-to-ne correspondence $x(t) \leftrightarrow x_0$.
Fig.\ref{fig:4}b shows the appearance of the first caustic (a
particle overtakes another). At Fig.\ref{fig:4}c the folds are
more pronounced and clearly visible. Finally at large times
(Fig.\ref{fig:4}.d) one can evidently observe the effect of the
clustering of particles.

Let us summarize the peculiarities of the evolution of the
distribution of inertial particles that distinguish them from
smooth compressible flows: 1) Infinite moments of density and
inter-particle distance may appear non-analytically at $t=+0$;
2) Average distance between particles grows exponentially; 3)
Moments of density in the Eulerian reference frame grow with the
rates  not reducible to those of distance moments in the
Lagrangian frame. The work was supported by the Israeli Science
Foundation, the EPSRC and the Royal Society.

\end{document}